\documentstyle[12pt,cite,epsf,epsfig]{article}
\newcommand{\newc}{\newcommand}
\newc{\tev}{\,{\rm TeV}}
\newc{\gev}{\,{\rm GeV}}
\newc{\sgn}{\mr{sgn}\,}
\newc{\ra}{\rightarrow}
\newc{\rpv}{$\mathrm{\not\!R_p}$}
\newc{\met}{$\not\!\!E_T$}
\newc{\rp}{$\mathrm{R_p}$}
\newc{\real}{\mathcal{R}e}
\newc{\alsm}{{\displaystyle \sum_{\alpha=1,2}}}
\newc{\besm}{{\displaystyle \sum_{\beta=1,2}}}
\newc{\al}{\alpha}
\newc{\ga}{\gamma}
\newc{\de}{\delta}
\newc{\cw}{\cos\theta_w}
\newc{\ssw}{\sin^2\theta_w}
\newc{\ccw}{\cos^2\theta_w}
\newc{\cbe}{\cos\beta}
\newc{\sbe}{\sin\beta}
\newc{\sh}{\hat{s}}
\newc{\sa}{\sin\al}
\newc{\ca}{\cos\al}
\newc{\bv}{$\mathrm{\not\!B}$}
\newc{\lv}{$\mathrm{\not\!L}$}
\newc{\ie}{{\it i.e.\/}\ }
\newc{\lam}{\lambda}
\newc{\cht}{\tilde{\chi}}
\newc{\upt}{\tilde{u}}
\newc{\elt}{\tilde{\ell}}
\newc{\hgt}{\tilde{H}}
\newc{\nut}{\tilde{\nu}}
\newc{\dnt}{\tilde{d}}
\newc{\psb}{\bar{\psi}}
\newc{\rtt}{\sqrt{2}}
\newc{\mut}{\tilde{\mu}}
\newc{\mr}{\mathrm}
\newc{\bath}{\bar{\theta}}
\newc{\tht}{\theta}
\newc{\JC}{{\bf J}}
\newc{\lra}{\longrightarrow}
\newc{\eg}{{\it e.g.\,}}
\newc{\barr}{\begin{array}}
\newc{\earr}{\end{array}}
\newc{\dis}{\displaystyle}
\newc{\beq}{\begin{equation}}
\newc{\eeq}{\end{equation}}
\newc{\me}{\mathcal{M}}
\newc{\dbm}{\partial_\mu}
\newc{\sgm}{\sigma_\mu}

\def\ra{\rightarrow}
\def\dis{\displaystyle}
\catcode`@=11 
\def \gsim{\mathrel{\mathpalette\@versim>}}
\def \lsim{\mathrel{\mathpalette\@versim<}}
\def \@versim#1#2{\lower0.4ex\vbox{\baselineskip\z@skip\lineskip\z@skip
     \lineskiplimit\z@\ialign{$\m@th#1\hfil##\hfil$%
     \crcr#2\crcr\sim\crcr}}}
\catcode`@=12 
\def\gev{\: \rm GeV}

\begin{document}
\setcounter{page}{0}
\renewcommand{\thefootnote}{\fnsymbol{footnote}}
\thispagestyle{empty}

\begin{titlepage}
\vspace{-2cm}
\begin{flushright}
BU-HEPP-07-10\\[2ex]
\end{flushright}
\vspace{+2cm}

\begin{center}
 {\Large{\bf W-pair Production in Unparticle Physics}
}\\
\vskip 0.6 cm
{\bf Swapan Majhi}
\footnote{E-mail: \ swapan\_majhi@baylor.edu}
        \vskip 1cm

{\it Department of Physics, Baylor University, Waco, TX 76706, U.S.A.
          }
\\
\vskip 0.2 cm
\end{center}
\setcounter{footnote}{0}
\begin{abstract}\noindent
 We consider the $W$-pair production for both $e^+e^-$ and hadron colliders in the context of unparticle physics associated with the scale invariant sector proposed  by Georgi. We have shown that the unparticle contributions are quite comparable with standard model (SM) specially for low values of non-integral scaling dimension $ (d_{\cal U}) $  and hence it is worthwhile to explore in current and future colliders.
\end{abstract}
\end{titlepage}
\setcounter{footnote}{0}
\renewcommand{\thefootnote}{\arabic{footnote}}

\setcounter{page}{1}
\pagestyle{plain}
\advance \parskip by 10pt

\section{Introduction}

      The novel idea of scale invariance plays a crucial role in both physics and mathematics.
For example, phase transition and critical phenomena are scale invariant at critical temperature
since all other length scale are considered as fluctuations which are equally important as well. In particle physics, scale invariance also plays an important role. Conformal invariance, in string theory is one of the fundamental property. This symmetry is broken in renormalisable field theories either explicitly  by some mass parameter in the theory or implicitly by quantum loop effects\cite{coleman}. In low energy particle physics, we observed different particles (elementary or composite) with different masses which is the consequence of such broken symmetry. Nonetheless, there could be a different sector of theory in four space-time dimensions which is exactly scale invariant and very weakly interacting with our low energy world (i.e. with standard model (SM) particles). 

           Recently, Georgi\cite{georgi} inspired by the Banks-Zaks theory\cite{BZ}, proposed a scale invariant sector (BZ) with non-trivial infrared fixed point. In such scale invariant sector, there are no particles since there is no particle state with a definite nonzero mass. Such sector is made of ``unparticles''. This BZ sector interacts with SM sector through exchange a very heavy (unspecified) particles with a large mass scale $M_{\cal U}$. Below this scale $M_{\cal U}$,  two sector interacts like a non-renormalisable theory which suppressed by powers of $M_{\cal U}$. On the other hand, scale invariance in the BZ sector emerge at an energy scale $\Lambda_{\cal U}$. The renormalisable couplings of the BZ field induce dimensional transmutation\cite{coleman} and the scale invariant unparticle emerge below an energy scale $\Lambda_{\cal U}$. Below the scale $\Lambda_{\cal U}$, the BZ sector is matched onto the unparticle operator and the nonrenormalisable interaction is matched onto a new set of interactions between SM and the unparticle fields with small coefficients. Such theory has very interesting phenomenological consequences\cite{georgi,LZ,CH12,Ding_Yan,Y_Liao,Aliev,Catterall,
Li,Lu,Steph,Fox,Greiner,Davou,CGM,XG,MR,Zhou,FFRS,Bander,Rizzo,GN,
Zwicky,Kikuchi,M_Giri,Huang,Krasnikov,Lenz,C_Ghosh,Zhang,Nakayama,
Deshpande,DEQ,Neubert,Hannestad,Das,Bhattacharyya,Majumdar,Alan-Pak,FD,Luis,Thomas}.

      In this article, we concentrated on $W$-pair production in both $e^+e^-$ and Hadron collider. We consider only two types of effective operators - scalar unparticle $O_{\cal U}$ and the spin-2 unparticle $O_{\cal U}^{\mu\,\nu}$. Feynman rules for these operators (which will couple to SM particles) are given in \cite{Cheung2}. For the sake of completeness, we are writing down the common effective interactions which satisfy the standard model gauge symmetry.
\beq
\lambda_0 {1 \over \Lambda_{\cal U}^{d_{\cal U}-1}} \bar{f} f O_{\cal U} \;,\hspace*{1cm}
\lambda_0 {1 \over \Lambda_{\cal U}^{d_{\cal U}}} G_{\alpha \beta} G^{\alpha \beta} O_{\cal U} 
\label{spin0}
\eeq
\beq
-{1 \over 4}\lambda_2 {1 \over \Lambda_{\cal U}^{d_{\cal U}}} \bar{\psi} i 
\Big(\gamma_{\mu} \stackrel{\leftrightarrow}D_{\nu} +\gamma_{\nu} \stackrel{\leftrightarrow}D_{\mu} \Big) \psi \,
O_{\cal U}^{\mu\,\nu},\;
\lambda_2 {1 \over \Lambda_{\cal U}^{d_{\cal U}}} G_{\mu \alpha} G^{\alpha}_{\nu} O_{\cal U}^{\mu\,\nu}
\label{spin2}
\eeq
where the covariant derivative $D_{\mu} = \partial_{\mu} + ig{\tau^a \over 2}W^a_{\mu} + i g^{\prime} {Y \over 2} B_{\mu}$, $G^{\alpha \beta}$ denotes the gauge field (gluon, photon, weak gauge bosons). $\psi$ stands for SM fermion doublet or singlet and $\lambda_i$ is the dimensionless effective couplings of the scalar ($i = 0$)  and tensor ($i = 2$) unparticle operators. The $W$-pair will be produced through both spin-0 and  spin-2 unparticle exchange (as given in eqns.(\ref{spin0},\ref{spin2})) in both $e^+e^-$ as well as hadron colliders.

This paper is organised in the following way. In section \ref{e+e-}, we discuss the total cross section and the differential distribution in the case of $e^+e^-$ collider and in section \ref{hadron}, we discuss the total cross section and differential distributions in the case of hadron collider. Finally we conclude in section \ref{conclusion}.

\section{$W$-pair production at $e^+\,e^-$ collisions}
\label{e+e-}

\begin{figure}[!h]
\centerline{
\hspace*{-2.0cm}
\epsfxsize=16cm\epsfysize=16.0cm
                     \epsfbox{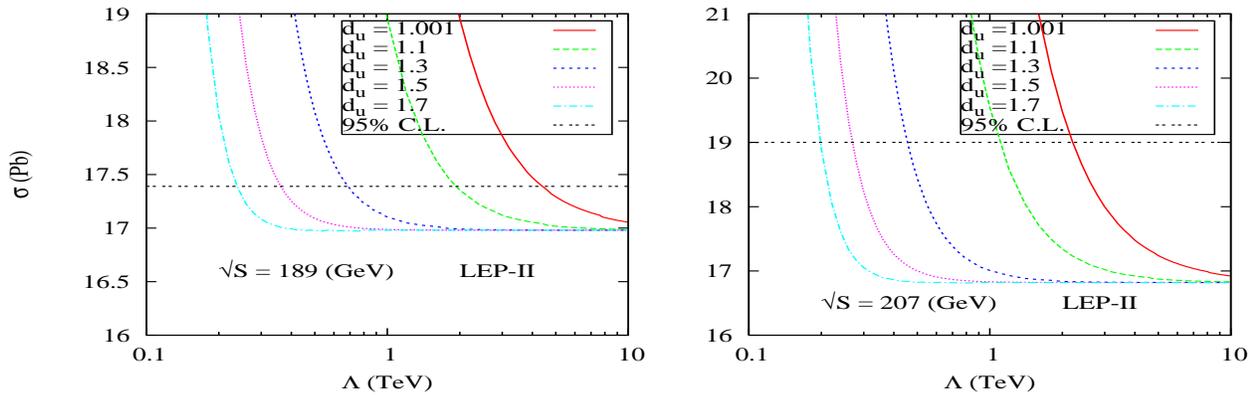}
}
\vspace*{-10.5cm}
\caption{\em The $W$-pair production cross section as a function of new scale $\Lambda_{\cal U}$ with various values of $d_{\cal U}$ and $\lambda_i = 1 (i=0,2)$. 
        }
\label{fig:lambda_cs}
\end{figure}

         The differential cross section for the process $e^+(p_1) e^-(p_2) \rightarrow W^+(p_3) W^-(p_4)$ is given by
\beq
{d\sigma \over d\Omega} = {1 \over 64 \pi^2 s} {|p_f| \over |p_i|} \sum_{A,B} \overline{|{\cal M}|^2_{AB}} \hspace{0.5cm} (A,B = U,\gamma,Z,t)
\label{eeWW_cs}
\eeq
where $U$ represents the spin-2 unparticle exchange diagram and $t$ represents the $t$-channel diagram. $p_i$ and $p_f$ are the three momentum of the initial and final state particle respectively. All the matrix element square are given in the Appendix. By integrating eqn.(\ref{eeWW_cs}) over angular co-ordinates, one can get the total cross section.
For numerical evaluation, we use the following input parameters:
\begin{eqnarray}
\alpha &=& {1 \over 137.04}   \hspace*{1.5cm} \sin^2\theta_W = 0.23
\nonumber\\
m_W &=& 80.403                \hspace*{1.5cm} m_Z = 91.1876
\nonumber\\
\Gamma_Z &=& 2.4952.
\nonumber
\end{eqnarray}
 We have plotted the $W$-pair production cross section versus the new scale $\Lambda_{\cal U}$ for two LEP-II energies (as shown in Fig.\ref{fig:lambda_cs}). From the  Fig.\ref{fig:lambda_cs} it is easy to read the upper bound on $\Lambda_{\cal U}$ for various values of $d_{\cal U}$ on the basis of SM measured value of cross section\cite{Opal} at $95\%$ C.L.(horizontal lines). By combining these two results, we put the upper bound for $\Lambda_{\cal U}$ at $95\%$ C.L. which is given in Table 1.

\begin{center}
\begin{tabular}{|c|c|}
\hline
{$d_{\cal U}$}  & {$\Lambda_{\cal U}$ (TeV)} \\
\hline
1.001 & 4.23\\
1.1 & 1.88 \\
1.3 & 0.68 \\
1.5 & 0.37 \\
1.7 & 0.25\\
\hline
\end{tabular}
\end{center}
\centerline{Table.1 Limits on $\Lambda_{\cal U}$ from the LEP-II data\cite{Opal} at 95$\%$ C.L. }


\begin{figure}[!h]
\centerline{
\hspace*{-2.0cm}
\epsfxsize=16cm\epsfysize=16.0cm
                     \epsfbox{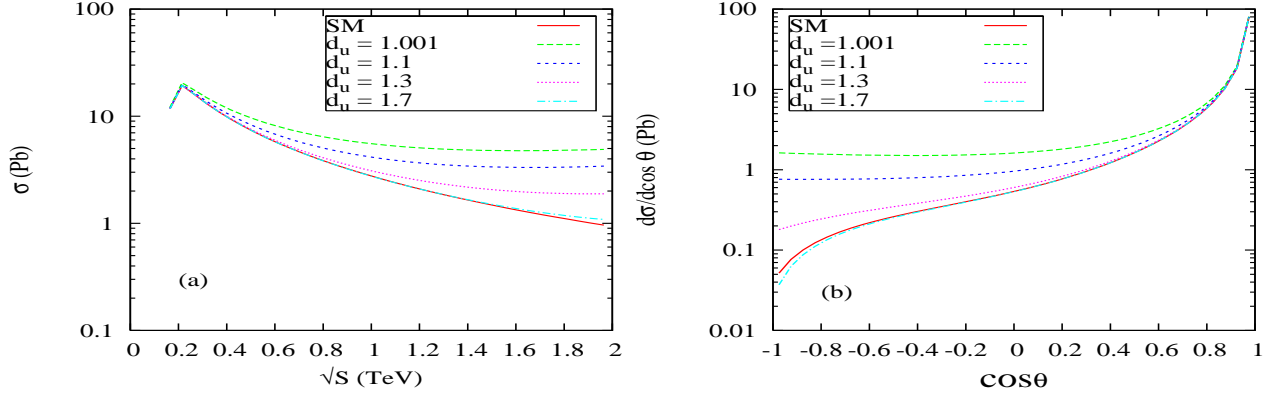}
}
\vspace*{-10.5cm}
\caption{\em (a)The $W$-pair production cross section as a function of center of mass energy $\sqrt{S}$ and $\Lambda_{\cal U} = 1$ TeV, $\lambda_i = 1 (i=0,2)$ (b) Angular distribution at $\sqrt{S} = 0.5$ TeV with $\Lambda_{\cal U} = 1$ TeV, $\lambda_i = 1 (i=0,2)$ All the curves are due to both contribution SM plus unparticle physics (including interference) except labeled by ``SM''. The label ``SM'' implies only SM cross section.
        }
\label{fig:tot_cs}
\end{figure}
      In Fig.\ref{fig:tot_cs}a, we have plotted the total cross section as a function of center of mass energy for various values of $d_{\cal U}$. From the figure, it is clear that for $\sqrt{S} > 500$ GeV, the spin-2 unparticle exchange contribution is significant compared to SM. As $d_{\cal U} \rightarrow 1$, the cross section due to unparticle contribution dominates over the SM. 

      As mentioned in paper\cite{georgi2}, the unparticle propagator has an extra phase exp$(-i \pi d_{\cal U})$ which can interfere with real photon propagator as well as both real and imaginary part of the $Z$-boson propagator. The imaginary contribution is quite small compared to real part due to the fact that it is proportional to $Z$-width. For example, $d_{\cal U} = 1.5$, only imaginary part will contribute to the cross section which is quite small compared to the real contribution. In Fig.\ref{fig:tot_cs}b, we have shown the angular distribution of the $W$-pair production. Since $W$-pair can only be produced through $s$-channel in unparticle case, the different operator structure will not matter in the experimental determination of cross section (see for example Fig.\ref{fig:tot_cs}b). As mentioned in the above, the new physics contribution starts to show up  for $d_{\cal U} < 1.3$.

\section{$W$-pair production at Hadron collider}
\label{hadron}
    
       The other process of interest to us in $W$-pair production is proton-proton (anti) collision, $P + P(\bar{P}) \rightarrow W^+ + W^- + X$, where $X$ implies a sum over all unobserved additional debris. In this case both spin-0 and spin-2 unparticle exchange  diagram will contribute. At the parton level, the processes are given below (through the effective operators as given in eqns.(\ref{spin0},\ref{spin2})).
\beq
q(p_1) + \bar{q}(p_2) \rightarrow W^+(p_3) + W^-(p_4) 
\label{fermion}
\eeq
\beq
g(p_1) + g(p_2) \rightarrow W^+(p_3) + W^-(p_4) 
\label{gluon}
\eeq

     The hadronic cross section is defined by convolution of partonic cross section with parton distribution functions and can be written as
\beq
d\sigma^{H_1 H_2} = \sum_{i,j} \int dx_1 dx_2  f_{i/H_1}(x_1) f_{j/H_2}(x_2) d\hat{\sigma}_{ij}(\hat{s},\hat{t}_1,\hat{u}_1) + (x_1 \leftrightarrow x_2)
\label{hadronic_cs}
\eeq
where $f_{i/H}(x_i)$ is the probability (usually called parton distribution function, PDF) of emitting a $i^{th}$-parton with a momentum fraction $x_i$ from a hadron $H$. The standard partonic mandelstum variables (defined in Appendix) $\hat{s},~\hat{t}_1,~\hat{u}_1$ are related to the hadronic variables ($S,~T_1,~U_1$) as $\hat{s} = x_1 x_2 S$, $\hat{t}_1 = x_1 T_1$ and $\hat{u}_1 = x_2 U_1$. The $\hat{\sigma}_{ij}$ is the partonic cross section. The above eqn.(\ref{hadronic_cs}) can be written as
\beq
S^2 {d^2\sigma^{H_1 H_2} \over dT_1 dU_1} = \sum_{i,j} \int_{{x_1}_{min}}^1 {dx_1 \over x_1} f_{i/H_1}(x_1) \int_{{x_2}_{min}}^1 {dx_2 \over x_2} f_{j/H_2}(x_2)\, \hat{s}^2 {d^2\hat{\sigma}_{ij} \over d\hat{t}_1d\hat{u}_1} (\hat{s},\hat{t}_1,\hat{u}_1) + (x_1 \leftrightarrow x_2)
\label{hadronic_cs2}
\eeq
where ${x_1}_{min},{x_2}_{min}$ are determined by the kinematic conditions
\begin{eqnarray}
\hat{s}+\hat{t}_1+\hat{u}_1 &=& 0
\hspace*{1cm}
x_1 x_2 S + x_1 T_1 + x_2 U_1 = 0 \hspace*{1cm} 0\le x_1 \le 1, ~ 0\le x_2 \le 1
\nonumber\\
{x_1}_{min} &=& {-U_1 \over S+T_1} \hspace*{1cm} {x_2}_{min} = {-x_1 T_1 \over x_1 S+U_1}.
\end{eqnarray}
\begin{figure}[!h]
\hspace*{-2em}
\centerline{
\epsfxsize=16cm\epsfysize=16.0cm
                     \epsfbox{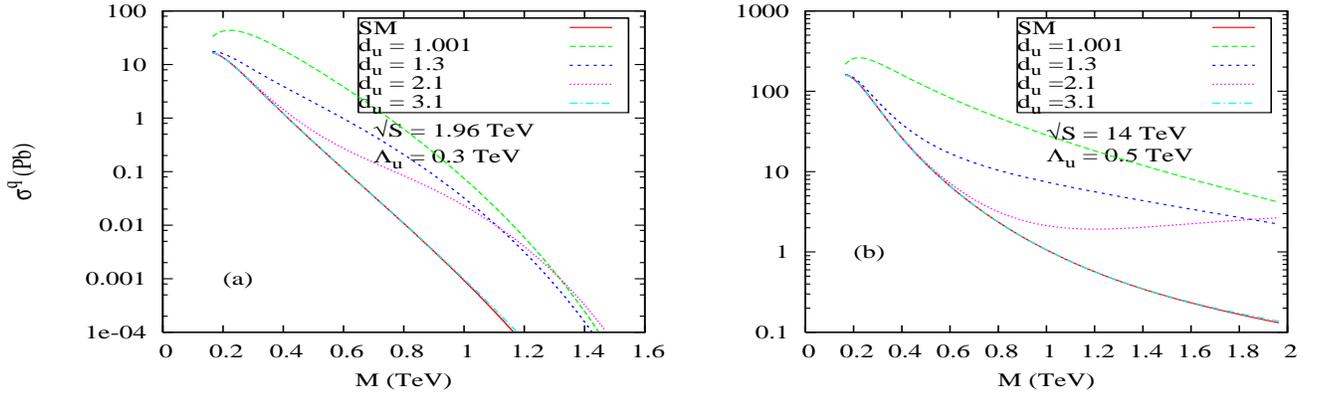}
}
\vspace*{-10.5cm}
\caption{\em $W$-pair production cross section versus invariant mass of the $W$-pair for the $q\bar{q}$-initiated process at Tevatron (LHC) and  $\lambda_2 = 1$.  The label ``SM'' implies only SM cross section.
        }
\label{fig:tot_cs_tev_lhc}
\end{figure}
\begin{figure}[!h]
\hspace*{-2em}
\centerline{
\epsfxsize=16cm\epsfysize=16.0cm
                     \epsfbox{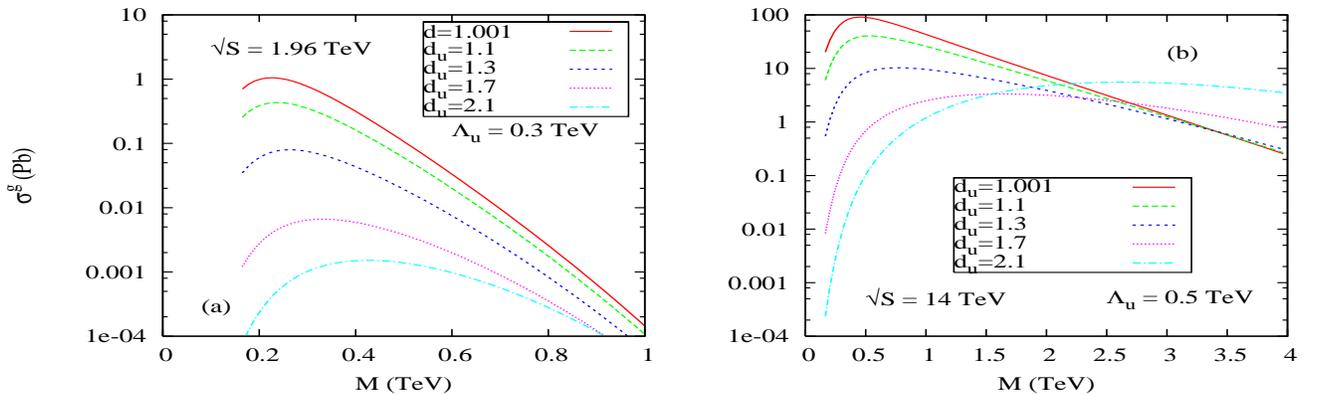}
}
\vspace*{-10.5cm}
\caption{\em $W$-pair production cross section versus invariant mass of the $W$-pair for the $gg$-initiated process at Tevatron (LHC) and  $\lambda_2 = 1$ and $\lambda_0 = 1$.
        }
\label{fig:tot_cs_gg}
\end{figure}
For our purpose, the double differential partonic cross section $\hat{s}^2 {d^2\hat{\sigma}_{ij} \over d\hat{t}_1d\hat{u}_1} (i,j = q, \bar{q} ~\&~ i,j = g,g)$ can be calculated from matrix element square given in Appendix. For numerical computation, we use CTEQ-6L1 parton distributions\cite{CTEQ6} with scale choice $Q^2 = \hat{s}$ as a factorisation scale.

\begin{figure}[!h]
\hspace*{-2em}
\centerline{
\epsfxsize=16cm\epsfysize=16.0cm
                     \epsfbox{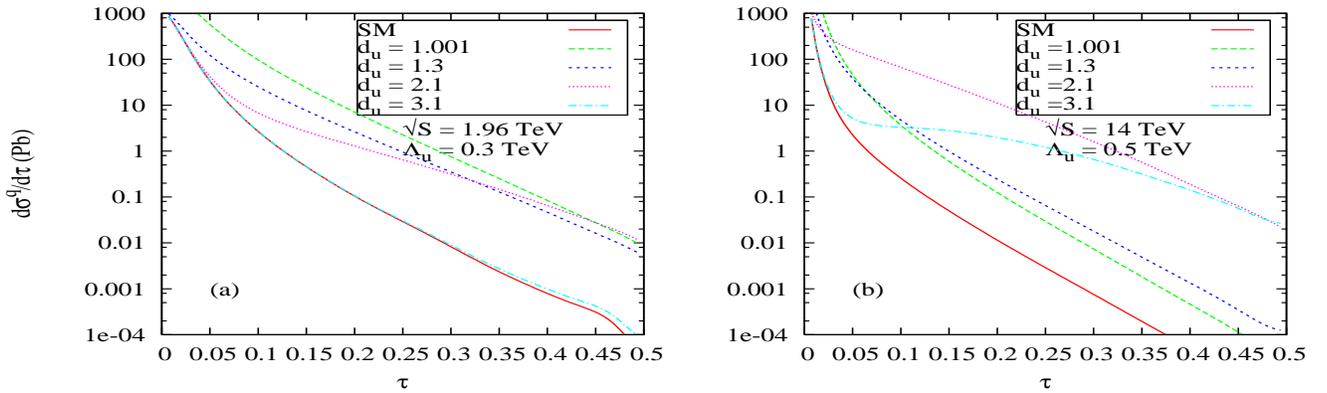}
}
\vspace*{-10.5cm}
\caption{\em $\tau (= \hat{s}/S)$-differential distribution  for the $q\bar{q}$-initiated process at Tevatron (LHC) and $\lambda_2 = 1$.  The label ``SM'' implies only SM cross section.
        }
\label{fig:tau_dist_lhc_tev}
\end{figure}

\begin{figure}[!h]
\hspace*{-2em}
\centerline{
\epsfxsize=16cm\epsfysize=16.0cm
                     \epsfbox{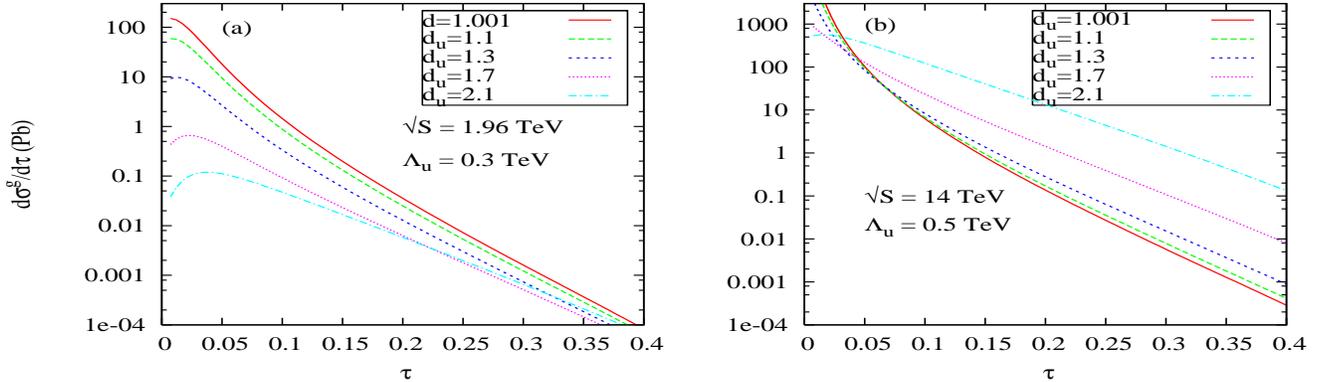}
}
\vspace*{-10.5cm}
\caption{\em $\tau$-differential distribution for the $gg$-initiated process at center of mass energy $\sqrt{S} = 1.96(14)$ TeV for Tevatron (LHC) and $\Lambda_{\cal U} = 0.3(0.5)$ TeV, $\lambda_2 = 1$ and $\lambda_0 = 1$. The label ``SM'' implies only SM cross section.
        }
\label{fig:tau_dist_gg}
\end{figure}

In Figs.(\ref{fig:tot_cs_tev_lhc}, \ref{fig:tot_cs_gg}) we have plotted the total cross section versus invariant mass of $W$-pair for both $q\bar{q}$ and $gg$-initiated processes for various values of $d_{\cal U}$. At LHC, $gg$ initiated process dominates over the $q\bar{q}$ process. That is mainly because gluon flux is larger than the $q\bar{q}$ flux. Whereas for Tevatron, it reverses the situation due to low center of mass energy and hence large $x_1\;,x_2$ dominated by the $q\bar{q}$ process. In spite of that the cross section of $q\bar{q}$-initiated process is large due to the presence of scalar interaction in both tevatron as well as LHC. This is true for rest of the analysis. We have also calculate the $\tau (= \hat{s}/S)$-distribution for the above mentioned processes as shown in Figs.(\ref{fig:tau_dist_lhc_tev},\ref{fig:tau_dist_gg}). From the figures it is clear that the differential cross section is better than the total cross section  for visibility study. This is due to the fact that in the total cross section we are integrating over the phase space as well as the parton momentum fractions $x_1$ and $x_2$.


\begin{figure}[!h]
\hspace*{-3em}
\centerline{
\epsfxsize=16cm\epsfysize=16.0cm
                     \epsfbox{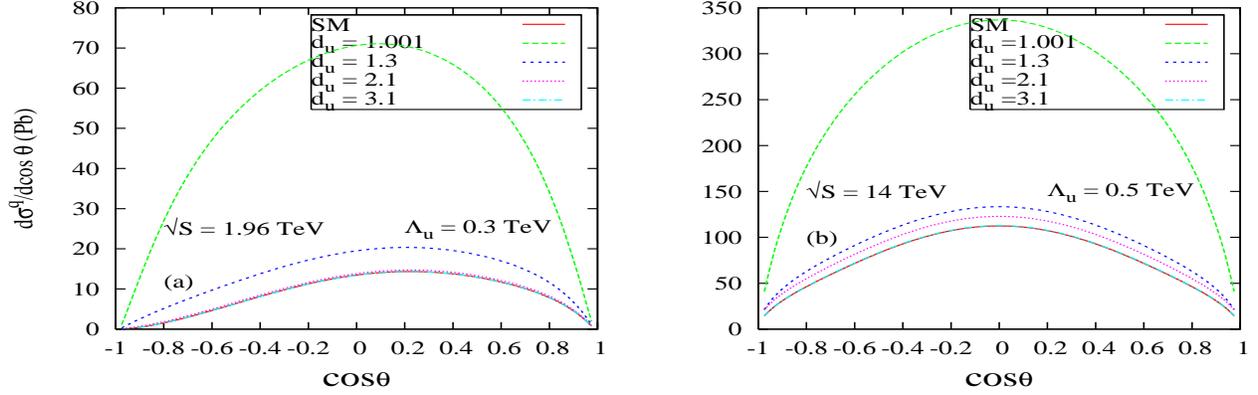}
}
\vspace*{-10.5cm}
\caption{\em Angular differential distribution of the $W$-pair production for $q\bar{q}$-initiated process at Tevatron (LHC) and  $\lambda_2 = 1$. The label ``SM'' implies only SM cross section.
        }
\label{fig:cth_dist_lhc}
\end{figure}

\begin{figure}[!h]
\hspace*{-2em}
\centerline{
\epsfxsize=16cm\epsfysize=16.0cm
                     \epsfbox{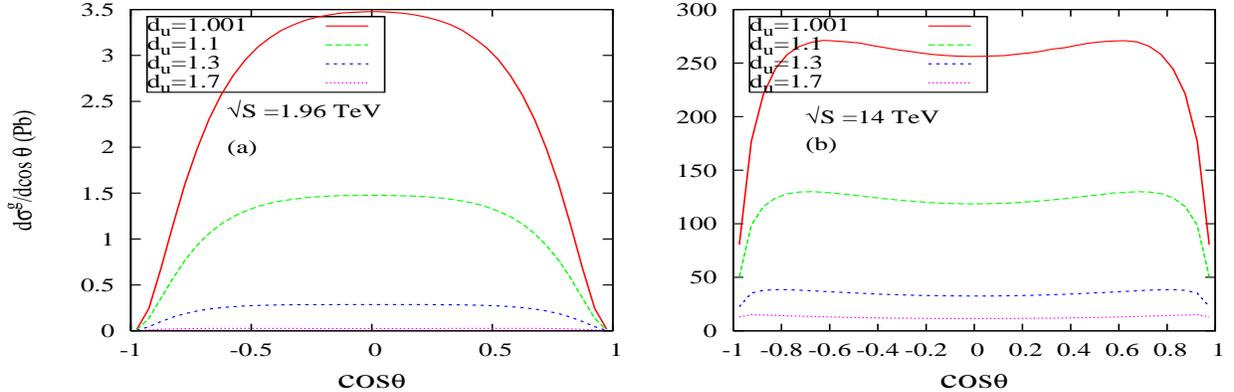}
}
\vspace*{-10.5cm}
\caption{\em Angular differential distribution  of the $W$-pair production for $gg$-initiated process at center of mass energy $\sqrt{S} = 1.96(14)$ TeV for Tevatron (LHC) and $\Lambda_{\cal U} = 0.3(0.5)$ TeV, $\lambda_2 = 1$ and $\lambda_0 = 1$. The label ``SM'' implies only SM cross section.
        }
\label{fig:cth_dist_gg}
\end{figure}

   In Figs.(\ref{fig:cth_dist_lhc},\ref{fig:cth_dist_gg}), we have plotted the angular distribution for both machines. Here $\theta$ is the parton rest frame scattering angle. To get the angular distribution in hadron frame, it has been boosted back to the hadron rest frame. For $q\bar{q}$-initiated process, the angular distribution is not symmetric at Tevatron due to the fact that the parton distribution functions are not symmetric under interchange of $x_1$ and $x_2$ whereas for LHC it is symmetric. At LHC, there is a small dip at the central region because the spin-2 unparticle exchange dominates over the scalar one for the $gg$-initiated process.

\begin{figure}[!h]
\centerline{
\epsfxsize=16cm\epsfysize=16.0cm
                     \epsfbox{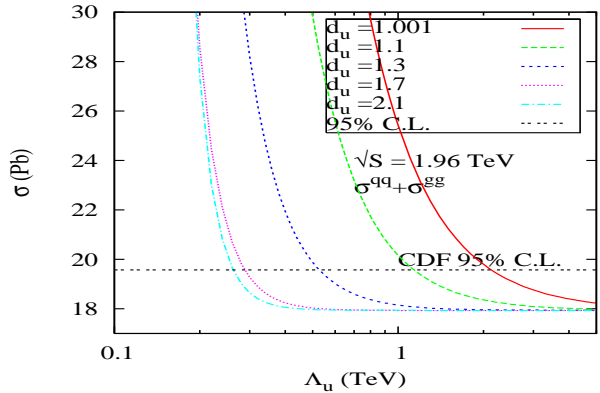}
}
\vspace*{-10.5cm}
\caption{\em The $W$-pair production cross section as a function of new scale $\Lambda_{\cal U}$ with various values of $d_{\cal U}$ with $\lambda_2 = 1$ and $\lambda_0 = 1$.
        }
\label{fig:lm_cs_tev}
\end{figure}

     In Fig.\ref{fig:lm_cs_tev} we display the $W$-pair cross section as a function of $\Lambda_{\cal U}$ for various values of $d_{\cal U}$ for combined ($q\bar{q}$ and $gg$-initiated) processes. Using the new CDF prelimenary result (horizontal lines in Fig.\ref{fig:lm_cs_tev} at 95$\%$ C.L.) on $W$-pair production\cite{CDF}, we put limits on $\Lambda_{\cal U}$ as given in Table.2. For a given value of $d_{\cal U}$, the  amplitudes scale as $\lambda^2_0/{\Lambda_{\cal U}}^{2d_{\cal U}-1}$ for scalar $q\bar{q}$-initiated process and $\lambda^2_i/{\Lambda_{\cal U}}^{2d_{\cal U}}(i=0,2)$ for abovementioned rest of the processes. 


\begin{center}
\begin{tabular}{|c|c|}
\hline
{$d_{\cal U}$}  & {$\Lambda_{\cal U}$} {(TeV)} \\
\hline
1.001 & 2.14\\
1.1 &1.14\\
1.3 &0.53\\
1.7 &0.29\\
2.1 &0.26\\
\hline
\end{tabular}
\end{center}
\centerline{Table.2 Limits on $\Lambda_{\cal U}$ from the CDF data\cite{CDF} at 95$\%$ C.L. }


The stronger bound comes from the scalar coupling of the unparticle with $q\bar{q}$ due to the fact that the power suppression factor ${\Lambda_{\cal U}}$ is less (by one at the amplitude level) than the other couplings.  So for fixed $\lambda_i(i=0,2)= 1$ the limits increases as $d_{\cal U}$ decreases from 2.1 to 1. 

\section{Conclusion}
\label{conclusion}
           In this article, we have calculated $W$-pair production in context of both $e^+e^-$ and hadron collider for various values of non-integral dimension $d_{\cal U}$. We showed that the scalar coupling of unparticle with fermions is dominated over the other couplings. From the discussion of sections(\ref{e+e-},\ref{hadron}) we can conclude that for $d_{\cal U} \le 1.3$, it is possible to discover the existence of unparticle (if it exists) in current and future colliders. The current measurement of LEP-II data, we put bound on $\Lambda_{\cal U}$ for different values $d_{\cal U}$.  We also put bound on parameter space ($\Lambda_{\cal U},d_{\cal U}$) with a fixed couplings $\lambda_i = 1 (i=0,2)$ at Tevatron. The bounds are strongly dependent on its mass dimension $d_{\cal U}$. In $e^+e^-$ case, the bound on scale $\Lambda_{\cal U}$ could be as large as few TeV as $d_{\cal U}$ close to 1 but for hadronic case, the bound on $\Lambda_{\cal U}$ is not so large as $e^+e^-$ case because of parton smearing. For larger value of $d_{\cal U}$, the bounds get weaker by power-law (${\Lambda_{\cal U}}^{2-4d_{\cal U}}$ and ${\Lambda_{\cal U}}^{-4d_{\cal U}}$). In hadron machine, apart from the $q\bar{q}$-initiated process, $W$-pair can be produced from $gg$-initiated process which is not present in SM (tree level) through the new effective interactions given in eqns.(\ref{spin0},\ref{spin2}). This has a large effect compared to SM  (specially at LHC). Hence the LHC allows us to investigate the gluonic couplings of the unparticle in the $W^+W^-$ mode and may lead to the discovery of unparticle physics.

\section*{Acknowledgements}

This work is supported
by US DOE grant No. DE-FG02-05ER41399 and NATO grant No. PST-CLG. 980342.

\section{Appendix }

\begin{eqnarray}
s = (p_1+p_2)^2\; ;
\hspace{.5cm}
t = (p_1-p_3)^2\; ;
\hspace{.5cm}
u = (p_1-p_4)^2\; ;
\hspace{.5cm}
\nonumber \\
p_1^2 = 0 \; ; \hspace{.5cm} p_2^2 = 0 \; ; \hspace{.5cm}
p_3^2 = m_{W}^2 \; ; \hspace{.5cm} p_4^2 = m_{W}^2 \; ;
\end{eqnarray}
\begin{eqnarray}
|M^S_{\cal U}|^2 &=& |B|^2 s\Big[(s-2m_W^2)^2 + 2 m_W^4\Big]
\end{eqnarray}
\begin{eqnarray}
|M^T_{\cal U}|^2 &=& 8 |A^{\prime}|^2\Big[4 u t (t^2+u^2) + m_W^2 s (t+u)^2 + 6 m_W^2 s t u
		+ 6 m_W^6 s - 8 m_W^8	\Big]
\end{eqnarray}
\begin{eqnarray}
|M_{\gamma + Z}|^2 &=& 4\Big(|f_L|^2+|f_R|^2\Big) s^2 \Big[\Big({u t \over m_W^4}-1\Big)
\Big({1 \over 4} - {m_W^2 \over s} + {3m_W^4 \over s^2}\Big) + {s \over m_W^2} -4\Big]
\end{eqnarray}
\begin{eqnarray}
|M_{t}|^2 &=& g_t^2 \Big[\Big({u t \over m_W^4}-1\Big)
\Big({1 \over 4} + {m_W^4 \over t^2}\Big) + {s \over m_W^2} \Big]
\end{eqnarray}
\begin{eqnarray}
2Re\Big[M^T_{\cal U}M_{\gamma + Z}^{\dagger}\Big]  &=& 8 Re\Big[A^{\prime}\;(f_L+f_R)\Big] 
(t-u)	\Big[t^2+u^2 + 4m_W^2s -2m_W^4	\Big]
\end{eqnarray}
\begin{eqnarray}
2Re\Big[M^T_{\cal U}M_{t}^{\dagger}\Big]  &=& 4 Re\Big[A^{\prime} g_t\Big] 
 		\Big[2t^2 + 2t(s-3m_W^2) +(s^2-2m^2_Ws + 6m_W^4 - {2m_W^6 \over t})\Big]
\end{eqnarray}
\begin{eqnarray}
2Re\Big[M_{\gamma+Z}M_{t}^{\dagger}\Big]  &=& 4Re\Big[f_L g_t\Big] s
 		\Big[\Big({u t \over m_W^4}-1\Big)
\Big({1 \over 4} -{m_W^2 \over 2s }- {m_W^4 \over s t}\Big) + {s \over m_W^2} -2 
+{2m_W^2 \over t}		
\Big]
\end{eqnarray}
\begin{eqnarray}
|M_{gg}|^2 &=& 32|A^{\prime}|^2 \Big(t^4+u^4 -4m_W^2(t^3+u^3)+4m_W^4(2t^2+2u^2+tu)-12m_W^6
(t+u) + 10m_W^8\Big)
\nonumber\\
&+& 64|B^{\prime}|^2 \Big((t+u)^4-4m_W^2(t+u)^3 + 6m_W^4(t+u)^2-8m_W^6(t+u) +8m_W^8\Big)
\end{eqnarray}
\begin{eqnarray}
g_t &=& g \hspace*{1.5cm} (\rm{for\; lepton})
\nonumber\\
&=& g V_{pn} \hspace*{1cm} (V_{pn} \rm{\;is\; the\; CKM\; mixing\; matrix}\; p = (u,c,t) \; n = (d,s,b))
\end{eqnarray}
\begin{eqnarray}
f_L &=& {e^2 e_f \over s} + {g^2  g_L \over 2 (s-m_Z^2 + i m_Z \Gamma_Z)} 
\; ; \hspace{.5cm} g \sin\theta_W = e 
\nonumber\\
f_R &=& {e^2 e_f \over s} + {g^2  g_R \over 2 (s-m_Z^2 + i m_Z \Gamma_Z)}
\; ; \hspace{.5cm} m_W^2 = \cos^2\theta_W\; m_Z^2
\nonumber\\
g_L &=& C_v^f + C_A^f \; ; \hspace{.5cm} g_R = C_v^f - C_A^f 
\; ; \hspace{.5cm} C_v^f = T^f_3 - 2 Q_f \sin^2\theta_W \; ; \hspace{.5cm}
 C_A^f = T^f_3
\end{eqnarray}
\begin{eqnarray}
A^{\prime} = {\lambda^2_2  Z_{d_{\cal U}} \over 4 \Lambda^{4}_{\cal U}} 
\Bigg({-s\over 
\Lambda^{2}_{\cal U} }\Bigg)^{d_{\cal U} - 2}
\hspace*{1cm}
B = {4\lambda_0^2 Z_{d_{\cal U}} \over \Lambda^3_{\cal U}} \Bigg({-s \over\Lambda^2_{\cal U}}\Bigg)^{d_{\cal U}-2}
\hspace*{1cm}
B^{\prime} = {\lambda_0^2 Z_{d_{\cal U}} \over \Lambda^4_{\cal U}} \Bigg({-s \over\Lambda^2_{\cal U}}\Bigg)^{d_{\cal U}-2}
\end{eqnarray}

\end{document}